\documentclass[%
 reprint,
 amsmath,amssymb,
 aps,
]{revtex4-2}

\usepackage{graphicx}
\usepackage{dcolumn}
\usepackage{bm}
\usepackage{float}

\begin{document}

\preprint{APS/123-QED}

\title{Dispersion Control in Micromechanical Evanescent Optical Modulators}

\author{Karl Johnson}\email{k9johnso@ucsd.edu}
\author{Noah A. Rubin}
\author{Yeshaiahu Fainman}
\affiliation{Department of Electrical and Computer Engineering, University of California San Diego, La Jolla, CA, USA}
\author{John Hong}
\author{Tallis Chang}
\author{Sean C. Andrews}
\author{Jean Huang}
\author{Leilani Ferguson}
\author{Liam McCue}
\author{Edward Chan}
\author{Bing Wen}
\affiliation{
 Obsidian Sensors, San Diego, CA, USA
}%
\date{\today}

\begin{abstract}

Efficient, low-loss, and versatile optical modulators are a critical ingredient for practical integrated photonic systems. Modulators based on micro-electromechanical systems (MEMS) have unique advantages over more traditional thermal, electro-optic, or plasma dispersion modulators. In this work, we show that evanescent MEMS modulators (in which a dielectric slab is mechanically inserted into a waveguide's evanescent field) can exhibit anomalously dispersive modulation. That is, despite positive modulation of a waveguide mode's effective index, the modulator brings about a negative change in group index.
We experimentally demonstrate these unique capabilities using a novel MEMS actuator design. The new theory and results here reveal that evanescent MEMS modulators possess a capability for control of wavelength dispersion not accessible to nearly any other type of modulator. These new capabilities may enable on-chip integration of systems for various optical applications, including broadband switching, photonic true time delay, pulse shaping, or phase matching of nonlinear processes.

\end{abstract}

\maketitle


\section{Introduction}

\begin{figure}
    \centering
    \includegraphics{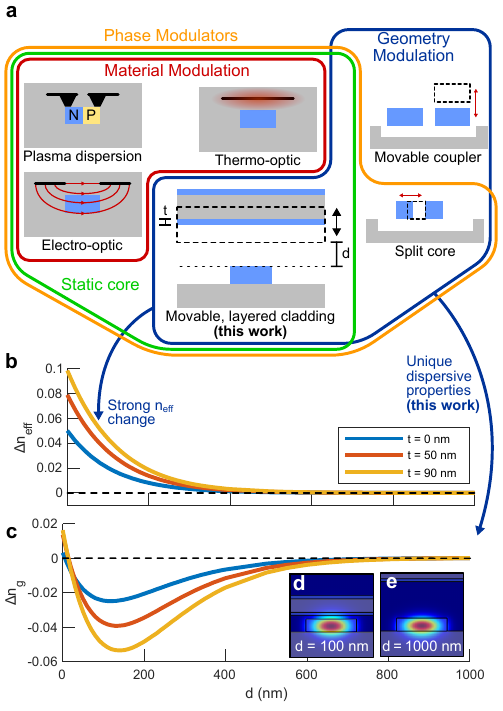}
    \caption{a) Cross sections of a few different modulator types used in integrated photonics. Movable cladding type modulators exhibit strong tunability of modal properties (characteristic of geometry modulation) while still benefiting from the advantages of devices with a static core. In this work, we utilize a device with a movable, multilayered cladding structure to highlight the unique dispersive properties of geometry modulation. b) Plot of the simulated effective index modulation as a function of the cladding-core separation $d$ for three different layer geometries, demonstrating large effective index modulations up to 0.1. c) Plot of the simulated group index modulation for the same three geometries, demonstrating the anomalous group index modulation effect emphasized in this work. Larger values of the thickness $t$ enhance the negative group index modulation, while also enabling zero and even positive group index modulation for small values of $d$. d,e) Electric field intensity of the waveguide mode in the $t = 90$ nm geometry for $d = $ 100 and 1000 nm.}
    \label{fig:big_picture}
\end{figure}

Over time, integrated photonic components have progressed from simple functions like passive routing, filtering, and (de)multiplexing \cite{marcatili1969dielectric, reed2005silicon} to highly complex active systems featuring high speed modulation \cite{alam2021net} and reconfigurability \cite{yao2023broadband}. Active systems necessarily require that the properties of waveguides and waveguide devices can be modulated - as such, substantial research effort is devoted to making integrated photonic modulators, and there are now a wide variety of mature modulator technologies. Some of these broad approaches are sketched in Fig. \ref{fig:big_picture}a. The simplest among these are perhaps thermo-optic modulators, which are very simple to implement and see widespread use \cite{haruna1981thermo, souza2018fourier, chung2019low}. However, thermal modulation often entails high power consumption, moderate speed, and limited modulation strength ($\Delta n_\text{eff}$), particularly for short wavelength applications where most waveguide materials have a very low thermo-optic coefficient \cite{johnson2022determination,chul2020chip,thomas1998frequency}. On the other hand, plasma dispersion modulators have seen widespread adoption for telecommunications owing to their high speed capability and CMOS compatibility \cite{reed2005silicon, siew2021review}. However, these modulators are unsuitable for broadband, low loss applications such as switching or tunable delay lines, as the injection of carriers to achieve modulation is fundamentally lossy and the modulation strength is still fairly weak \cite{soref1987electrooptical}. Electro-optic devices utilizing $\chi^{(2)}$ nonlinearity represent yet another option, capable of low loss modulation with extreme speed \cite{valdez2023100, hu2025integrated}. Nevertheless, limited CMOS compatibility and extremely weak modulation strength make these devices impractical for all but the most demanding applications in high-speed modulation where the large interaction lengths required for a $\pi$ phase shift are acceptable.

In the search for broadband, low loss, and high $\Delta n_\text{eff}$ modulators, the past 40 years have seen substantial efforts investigating the use of MEMS devices in integrated photonic systems \cite{chollet2016devices, ohkawa1989integrated, terui1981total}. Unlike all other modulators (where weak changes to the refractive index of the waveguide materials are the source of modulation), the ability for MEMS waveguide devices to dynamically change the geometry of the optical system yields a completely different set of capabilities. Many approaches to MEMS in integrated photonics elect to mechanically actuate the waveguide core itself; these devices naturally can be used to create switches with near-perfect extinction ratios and extremely strong modulation strengths \cite{akihama2011ultra, chatterjee2010nanomechanical, han2015large}. Unfortunately -- with some notable exceptions \cite{seok2016large} -- these devices typically require complex suspended structures in which the substrate under the waveguide layer is etched away, requiring large footprints and limiting compatibility with more conventional fabrication processes. Another approach is to keep all waveguide cores static and only actuate non-guiding dielectric structures that interact with the evanescent field surrounding the waveguide \cite{terui1981total, lukosz1991integrated, pliska1993electrostatically, dangel1998electro, gui1999fabrication, nielson2005integrated, pruessner2014optomechanical, hah2011mechanically}. While this limitation restricts this class of devices to phase modulation (as opposed to the direct switching or on/off amplitude modulation possible with movable waveguide structures), the $\Delta n_\text{eff}$ of these devices can still be very high. Furthermore, these evanescent MEMS devices can be fabricated on top of nearly any air-clad waveguide, agnostic to the waveguide material and without disrupting the substrate underneath.

There is substantial previous work presenting various designs and applications of such movable cladding-based MEMS waveguide devices. The simplest of these devices are variable attenuators in which the transmission through a waveguide can be decreased by the insertion of an absorbing material into the evanescent field \cite{gui1999fabrication}. Alternatively, insertion of a suitably high index dielectric in the evanescent field can frustrate the total internal reflection required for guiding (causing outcoupling), providing another method for variable attenuation \cite{terui1981total}. For low-loss switches and modulators, use of a movable MEMS slab with low absorption and a dielectric constant equal to or lower than the core material is required. Such modulators have been demonstrated in a variety of devices, including microphones \cite{lukosz1991integrated}, displacement sensors \cite{pruessner2014optomechanical}, switches \cite{dangel1998electro, nielson2005integrated}, and tunable filters \cite{hah2011mechanically}.

In this work, we show that these movable cladding devices exhibit surprising wavelength-dependent modulation characteristics. In particular, while insertion of a high index cladding into the waveguide's evanescent near-field yields a substantial (and expected) increase in the modal effective index (Fig. \ref{fig:big_picture}b), the group index \textit{decreases} (Fig. \ref{fig:big_picture}c) --- a less intuitive and markedly less expected effect. Furthermore, this distance-dependent group index shift exhibits a dramatic inflection point at close cladding-core distances, with the group index modulation rising sharply to zero or even positive values. While some authors have discussed the use of a MEMS modulator for group delay modulation \cite{horak2012coupled, lee2006tunable}, to our knowledge the anomalous behavior theoretically described and experimentally demonstrated here has not been previously identified.

This work begins by offering a simple explanation for this anomalous group index modulation and discussing how a layered refractive profile of the movable MEMS slab allows for this unique response to be engineered. To experimentally demonstrate this effect, we propose a MEMS actuator design that can achieve the nanometric control required for basic modulation. Furthermore, we show that control of the curvature of the MEMS slab can be used to further fine tune the modulation and also to lower the insertion loss of the modulator substantially. These properties are experimentally validated using experiments with the MEMS devices placed inside of add-drop ring resonators and Mach-Zehnder interferometers (MZIs). Using these measurements, we are able to perform extremely precise measurements of the phase, group delay, and loss properties of these MEMS modulators. Despite inconsistent actuation in our results due to limitations of the MEMS actuators as-fabricated, we show that the most fundamental optical effects sought here are extremely consistent for devices fabricated across a 4" wafer. While we demonstrate these general results using one specific MEMS actuator design, we emphasize that the results are independent of the choice of actuator, and are general to the interaction between a movable layered dielectric slab and a dielectric waveguide. This general capability of MEMS cladding modulators to achieve these unique dispersive effects in integrated photonic platforms may enable new possibilities for reconfigurability in on-chip switches, delay lines, and even for phase matching of nonlinear processes.

\section{Design, Theory, and Simulations}

\subsection{Anomalously dispersive modulation}

The key feature of the modulator discussed in this work is an adjustable cladding, implemented using a dielectric slab moved up and down above a waveguide core (Fig. \ref{fig:big_picture}a, bottom). When the slab-waveguide distance $d$ is small enough (typically $<\lambda/2$), the MEMS slab enters the evanescent field above the waveguide and the effective index of the waveguide mode is increased. For a simple (flat) modulator which has an unchanging cross section over an interaction length $L$, we can express the total phase modulation of the modulator at a given wavelength as 

\begin{equation}\label{eq:phase_neff}
    \Delta\phi(d) = L\Delta k  = 2\pi\frac{L}{\lambda_0}\Delta n_\text{eff}(d) 
\end{equation}

where $\Delta n_\text{eff}(d_0) = n_\text{eff}(d_0) -  n_\text{eff}(d \rightarrow\infty)$ is the effective index modulation as a function of distance.

To further analyze the optical properties of the cladding modulator, we performed finite difference frequency domain (FDFD) simulations for $\lambda = $ 1650 nm light on the device cross-section shown in Fig. \ref{fig:big_picture}a (bottom) which consists of a 350 x 1400 nm Si$_3$N$_4$ waveguide on flat SiO$_2$ substrate. The MEMS slab above the core features a multilayer dielectric structure with a low index SiON core ($n\approx1.5$, 360 nm thickness) surrounded (above and below) by a higher index SiN thin film ($n\approx1.9$). As shown in Fig. \ref{fig:big_picture}b, moving the high index slab into the waveguide's evanescent near-field yields a large increase in $n_\text{eff}$, which has been shown in several previous works featuring similar modulators \cite{lukosz1991integrated, pruessner2014optomechanical, dangel1998electro, nielson2005integrated, hah2011mechanically}. When the thickness of the SiN film is increased, the maximum $n_\text{eff}$ modulation increases. This is intuitive, given that the average material index of the dielectric slab has increased as well. While increasing this thickness is a reasonable way to increase the maximum $n_\text{eff}$ modulation, this film cannot be made arbitrarily thick. If made too thick, the nitride film will eventually begin to support its own well-confined slab mode which will couple light out of the main waveguide core, where it will scatter into free space at the end of the modulator.

In addition, the presence of the MEMS slab also affects the group index of the waveguide mode, which in contrast takes into account the variation of the effective index with wavelength. Fig. \ref{fig:big_picture}c shows the behavior of the group index ($n_g$) with distance $d$, the results of which are perhaps more surprising than those of $\Delta n_\text{eff}$. To aid in this discussion we also plot the mode profiles in Fig. \ref{fig:big_picture}d-e which show the mode for two $d$ values in the $t$ = 90 nm case. Starting on the right side of Fig. \ref{fig:big_picture}c ($d\rightarrow \infty$), as the MEMS slab begins to move into the waveguide field, $n_g$ first decreases. However, once the slab becomes close enough, the curve inverts and begins to increase again. In the $t$ = 50 and 90 nm cases, the $n_g$ modulation even crosses over 0 and becomes positive once the modulator reaches $d = 0$. This unique behavior is best understood by first recalling the common definition of $n_g$,

\begin{equation}
    n_g(\lambda_0) = n_\text{eff}(\lambda_0) - \lambda_0 \frac{dn_\text{eff}}{d\lambda}.
\end{equation}

The term $- \lambda_0 \frac{dn_\text{eff}}{d\lambda}$ is responsible for the initial negative trend of the $n_g$ modulation with $d$. This is fundamentally because longer wavelengths are less confined and have a longer evanescent tail above the waveguide. Given this, longer wavelengths experience modulation both earlier (larger $d$) and more strongly than shorter wavelengths. Consequently, modulation of the $\frac{dn_\text{eff}}{d\lambda}$ term is positive, and $- \lambda_0 \frac{dn_\text{eff}}{d\lambda}$ is negative. Meanwhile, as shown in Fig. \ref{fig:big_picture}b, modulation of the $n_\text{eff}(\lambda_0)$ is always positive. Competition between these two terms leads to the behavior seen in Fig. \ref{fig:big_picture}c - in Supplemental Section S2 we plot these two terms separately to illustrate this. At longer gaps, the second term dominates (because wavelengths longer than $\lambda_0$ begin to impact the derivative term before the $n_\text{eff}(\lambda_0)$ term is impacted), but at very close gaps the wavelength dependence is less strong, so the derivative term lessens and $n_\text{eff}(\lambda_0)$ dominates. By adjusting the thickness $t$ of the layer on the bottom of the slab, the maximum magnitude of the negative group index modulation as well as the value of $d$ at which the turning point occurs can be engineered.

Given the unique behavior of the group index, it is also sensible to examine the behavior of the group velocity dispersion (GVD) with varying $d$. Indeed, as plotted in Supplemental Section S3, the GVD also exhibits similar turning-point behavior, with both positive and negative signs of the modulation easily accessible. We leave further discussion of the GVD to Supplemental Section S3 as the short interaction length of the devices fabricated in this work prevented us from reliably experimentally observing this effect.

The plots in Fig. \ref{fig:big_picture}d-e also illustrate another important impact of this modulation on the properties of the waveguide mode. When the slab is lowered onto the waveguide, the center of the mode profile noticeably shifts upward into the MEMS slab. This mismatch in mode profile versus that of the standalone waveguide introduces the possibility of large coupling losses at each edge of the modulator structure. This must be accounted for in the design of the MEMS actuator, as we discuss further below.

\subsection{MEMS Implementation}

\begin{figure}
    \centering
    \includegraphics{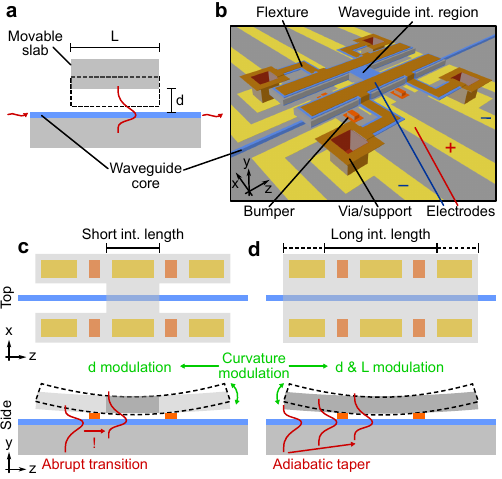}
    \caption{Main features of the MEMS actuator. a) Side profile illustrating the motion which must be implemented by the MEMS actuator: the cladding must be precisely moved in and out of the evanescent field extending vertically above the waveguide core. b) Three-dimensional rendering showing a representative example of the MEMS actuator design used in this work. c) Impact of slab curvature for short interaction length variants of the MEMS modulator: curvature modulation allows for fine tuning of $d$, but there is mode mismatch loss due to the abrupt transition between the bare waveguide mode and the mode in the modulated region. d) For long interaction length variants, modulation of the slab curvature tunes both $d$ and $L$. The slab curvature also serves as a low-loss taper between the bare waveguide mode and the modulated mode. }
    \label{fig:overview}
\end{figure}

The interesting optical phenomena discussed in the previous section can only be realized with a MEMS actuator capable of affecting both \textmu m-scale motions (to move the cladding completely out of the evanescent field) and nm-scale motions (for fine tuning of the modulation characteristics when the cladding is in the down state). Fig. \ref{fig:overview}a illustrates these operating requirements of the cladding modulator once again, this time using a side view which highlights the propagation direction of the light in the waveguide. To accomplish this, here we use electrostatically actuated MEMS devices of the design shown in Fig. \ref{fig:overview}b. As mentioned previously, the actuator design is not the main focus of this work, but we discuss it briefly here to illustrate how the general optical principles of the aforementioned evanescent interaction can be practically implemented. In our implementation, the dielectric slab is suspended above the waveguide using `crab-leg' style flexures \cite{fedder1994simulation}. As a voltage is applied between electrodes on the substrate and electrodes on the slab, an attractive electrostatic force actuates the slab in the y-direction (vertically). The metal electrodes are intentionally kept out of the interaction region above the waveguide by design to avoid excess absorption loss. In general the structure is designed with a 8 \textmu m wide (x-direction) keep-out region centered on the waveguide to minimize unintended optical effects from mechanical and electrical geometry. Vias between the substrate and the crab-legs provide vertical height above the substrate and electrical connection to the top electrodes.

An important aspect of electrostatic actuators is the phenomenon of `pull-in' caused by the nonlinear force-versus-distance of electrostatic attraction \cite{fedder1994simulation}. For a MEMS actuator starting in the `up' state, increasing the voltage will (at first) cause smooth motion of the MEMS actuator, but eventually this electrostatic force will overcome the spring, causing a sudden transition to a `pull-in' state in which the actuator is held tightly against the substrate. The most important practical consequence of this bistability effect here is that certain MEMS-waveguide distances $d$ are inaccessible, as there is no voltage for which a stable equilibrium exists at that position. To allow for precise positioning the MEMS slab in the pull-in state, we include bumpers on the substrate which are raised slightly (90 nm in this work) above the surface of the waveguide. In the case of a completely flat slab, it is this height of the bumpers above the waveguide top surface that determines the MEMS-waveguide distance $d$ in the pull-in state. However, in practice, both fabrication and actuation cause curvature of the MEMS slab which must be taken into account.

\subsection{Impact of MEMS Slab Curvature}

Our actuator design features a suspended structure with several layers of different materials. In such MEMS structures, there is often curvature of the suspended structure due to film strain \cite{chen2002modification}. This curvature may also be changed in the pull-in state by applying a nonuniform force across the MEMS structure. Here, we achieve this by continuing to increase the voltage of the center electrode past the point at which the slab has been ``pulled in'' and is resting on the aforementioned bumper structures (Fig. \ref{fig:overview}c). Since the bumpers are spaced on either side of this electrode, the attractive force selectively pulls the center towards the substrate, causing a convex curvature of the MEMS slab on the waveguide side. As the slab is thick, the spring constant of this secondary modulation is much higher than that of the crab-leg flexures, allowing for extremely fine voltage-tunability of $d$ in the down position.

For the remainder of this section, we consider only the optical properties of slab curvature, independent of how an actuator (or even which specific MEMS actuator) is used to achieve that curvature. There are two main cases to discuss when analyzing the impact of slab curvature, both short and long interaction length regimes. First, if the MEMS device has only a short modulation length (Fig. \ref{fig:overview}c), the curvature within the interaction region is fairly low and may be neglected, but the curvature of the rest of the structure is still important, as it is responsible for the fine-tuning of $d$ with voltage in the pull-in state. As will be shown later, optical properties change extremely rapidly with $d$ in the pull-in state, so nm-level control is essential for repeatable modulation. One important aspect of the short interaction length case is the abrupt transition from the bare waveguide to the modulated region. As previously mentioned, the modal profile in the modulated regions is shifted from that of the bare waveguide (Fig. \ref{fig:big_picture}d-e). Given this, the mode mismatch at this abrupt transition causes a mode mismatch, resulting in insertion loss. This source of insertion loss has been noted previously \cite{hah2011mechanically}. It should be noted that Fresnel reflection due to the $n_\text{eff}$ mismatch at this interface also contributes to the insertion loss; however, the mode profile mismatch is the dominant effect for the devices discussed in this work.

The second case to consider is when the interaction length is long enough that the MEMS slab curvature is not negligible within the waveguide interaction region (long length case, Fig. \ref{fig:overview}d). While it is possible for the slab's bottom surface to be either concave or convex, in the concave case there is very little interaction length other than at the ends of the modulator, so we will not examine that situation here. If there is a convex curvature in the pull-in state, the first observation is that changes in curvature not only yield small changes in $d$ in the center of the waveguide, but also changes in the effective interaction length $L$ between waveguide and slab. In other words, the effective interaction length and interaction distance \emph{both} vary simultaneously in these long interaction length devices. More formally, if we know the position-dependent MEMS-waveguide gap $d(z)$ and assume the evolution of the structure is adiabatic (does not scatter light into other modes), we can express the total phase shift of the curved structure at a given wavelength as

\begin{equation}\label{eq:phase_integral}
    \Delta\phi = \frac{2\pi}{\lambda_0}\int_0^L n_\text{eff}\big(d(z)\big)dz.
\end{equation}

Furthermore, in cases where the slab curvature is high enough that the ends of the MEMS slab are outside of the evanescent field, the curvature of the slab can act as a gradual transition between the bare waveguide and the modulated region, minimizing modal-mismatch. This allows for low loss modulation, so long as the structure is not laid so flat that the ends of the structure fall inside the evanescent field, re-introducing mode mismatch loss. 

\section{Fabrication}

\begin{figure}
    \centering
    \includegraphics{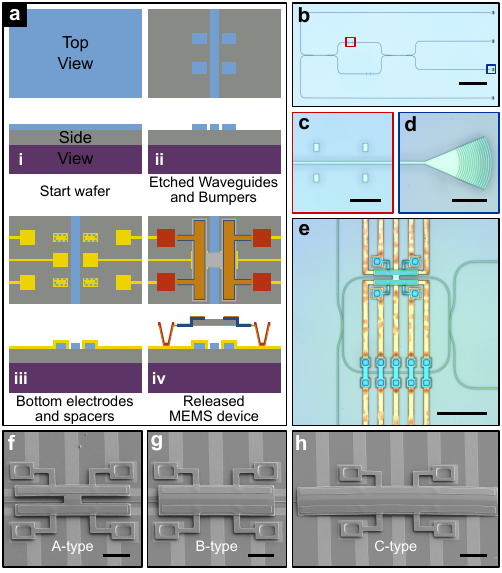}
    \caption{a) Overview of the fabrication process. b) Optical microscope image of an MZI structure prior to MEMS fabrication (scale bar 250 \textmu m). Close-up images of the highlighted regions show c) foundations for the MEMS bumpers and d) focused grating couplers (scale bar 20 \textmu m). e) Optical microscope image of a finished MEMS device on a ring resonator (scale bar 100 \textmu m). f-h) Scanning electron microscope images of three different finished MEMS device variants (scale bar 20 \textmu m).}
    \label{fig:fabrication}
\end{figure}

To experimentally demonstrate the unique properties of this evanescent interaction, we fabricated devices in the Nano3 cleanroom facility at UC San Diego (La Jolla, CA, USA). Fabrication was performed on 4'' wafers purchased from a commercial vendor (Applied Nanotools, Inc.) with 350 nm of low pressure chemical vapor deposition (LPCVD) Si$_3$N$_4$ and 4.5 \textmu m of buried SiO$_2$ pre-deposited on a 500 \textmu m Si substrate (Fig. \ref{fig:fabrication}a,i). Waveguide structures were then defined by electron beam lithography in a negative polymer photoresist (ma-N 2405), which were then etched using inductively coupled plasma reactive ion etching (ICP/RIE). The resist was stripped using oxygen plasma leaving finished waveguides (Fig. \ref{fig:fabrication}a, ii). Images of these bare waveguide structures are shown in \ref{fig:fabrication}b-d. Our design uses grating couplers for coupling light on and off chip (\ref{fig:fabrication}d), allowing for testing of the passive waveguide devices at this fabrication step before MEMS integration. Several devices were measured to verify acceptable yield of functional waveguide structures, and spectra from add-drop ring resonators were also used to verify the waveguide loss matched the expectation of 0.5-1 dB/cm from previous fabrications \cite{fedorov2025unambiguous}.

The first step of the MEMS fabrication used here is to deposit and etch the first metal layer, which serves two purposes. Primarily, this layer has an electrical function, including routing to devices, formation of the bottom electrodes of the electrostatic actuators, and contacts underneath the vias leading to the top electrodes. Additionally, this metal is used to form the `bumper' structures that define the MEMS-waveguide distance $d$ in the pull-in state. This is performed by depositing metal on top of bumper `foundations' included on the waveguide layer (Fig. \ref{fig:fabrication}c). As these foundations have the same height as the waveguides, the metal deposition merely serves to slightly increase the height of these foundations above the top surface of the waveguide. This self-referenced method allows for the bumper height dimension to be substantially more accurate and repeatable than if the bumpers were grown directly on the buried oxide.

The rest of the MEMS structure is then fabricated. The elevated geometry of the MEMS device is fabricated on top of a 2 \textmu m sacrificial layer, yielding a resting position of the MEMS slab that is at least 1.5 \textmu m above the waveguide, well outside of the evanescent field. The dielectric slab is composed of a 90 nm SiN / 360 nm SiON / 90 nm SiN layer stack to provide the desired optical properties and also balance the strain of the structure. The top layer of SiN is also used to form the crab-leg flexures. A second metal layer forming the top routing and electrodes is patterned on top of the SiN, before the device is released in oxygen plasma to suspend the MEMS structures.

An example of a complete device on top of a ring resonator is shown in \ref{fig:fabrication}e. Also visible at the bottom of this image are air bridges (utilizing a similar structure and sharing the same process steps as the MEMS actuators) which allow for metal routing and waveguides to cross without introducing optical loss. Across the 4'' wafer, numerous variants of the MEMS actuator were placed within different integrated optical test structures including all-pass ring resonators and MZIs. Scanning electron microscope images of three main MEMS device variants are shown in \ref{fig:fabrication}f-h. These three structures have physical lengths of 13, 87, and 162 \textmu m along the propagation direction of the waveguide, respectively. For brevity we refer to these respectively as A-, B-, and C-type modulators. The slab curvature of the released devices is generally well-controlled, however on this fabrication a select few devices (such as the modulator pictured in \ref{fig:fabrication}h) exhibit downward curvature (concave on the waveguide side).

\section{Optical Characterization and Results}

\begin{figure*}
    \centering
    \includegraphics{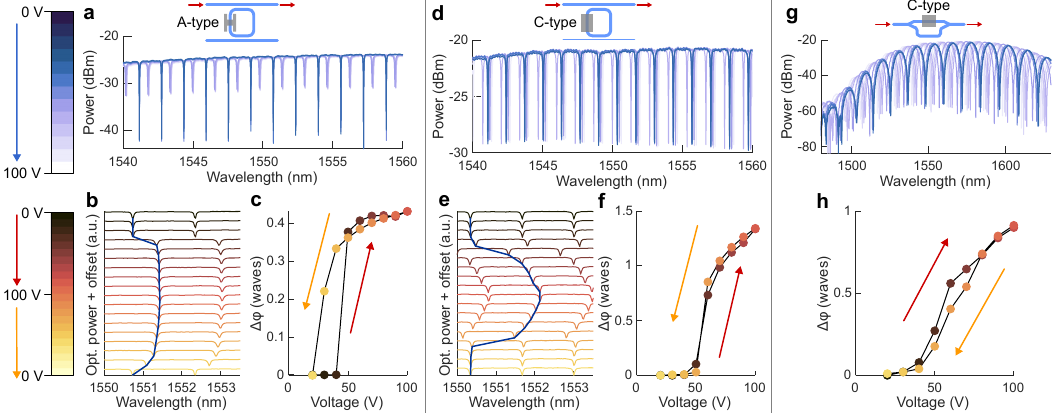}
    \caption{Optical spectra used to measure phase modulation. a)``Through'' spectrum of an add-drop ring resonator with an A-type MEMS modulator measured for voltages ranging from 0-100 V. b) Close-up of the spectra in a) plotted in linear scale and offset for clarity. The blue curve shows the tracked wavelength shift of a single resonance, which is proportional to the phase shift. c) Extracted phase shift as a function of voltage from the data in b). d-f) Same as a-c), but utilizing a longer interaction length (C-type) structure. Here, no changes in resonance extinction ratio are visible, signifying suppression of insertion loss by the curved structure. f) Transmission spectra versus voltage for an imbalanced MZI with a C-type MEMS modulator, and g) extracted phase shift.}
    \label{fig:spectra}
\end{figure*}

To characterize the optical properties of these modulators, we measure optical transmission spectra using a tunable laser (Santec TSL-550) and photodetector (Agilent 81635A and Thorlabs DET08CFC) while various voltages are applied to the MEMS structure (Keithley 2400). Diagrams and further details regarding the experimental setup are provided in Supplemental Section S5. To optically interface with the chip, we use a 250 \textmu m pitch fiber array configured to have a 12 degree angle of incidence in air, to match the grating coupler design. To electrically interface with the chip, we use 4 needle probes. While the fabricated structures have a total of 6 independently addressable bottom electrodes, due to the limited number of electrical probes we only apply a voltage to the center electrode on each of the modulator, with the top electrodes grounded and the remaining electrodes left disconnected. Use of only one third of the bottom electrodes leads to higher actuation voltages and less complete control over the curvature of the MEMS slab, but this level of control was sufficient for the purposes of the initial demonstrations shown in this work.

We note here that there exist complications in extracting and interpreting modulation results from devices (such as the long interaction length MEMS slabs with substantial curvature) where the effect of the modulator varies substantially along the propagation direction $z$. This topic is discussed in more depth in Supplemental Section S1, where we distinguish between spatially `instantaneous' modal properties (such as $n_\text{eff}$ and$ n_g$) and spatially `integrated' properties of an entire modulator (the phase modulation $\Delta\phi$ and the group delay modulation $\Delta\tau_g$). Supplemental Section S1 also discusses how these parameters are obtained from the measured optical spectra, using techniques similar to those used in previous works \cite{johnson2022determination, hah2011mechanically, belogolovskii2025large, gardes2009high}. Most results plotted in this discussion are the `integrated' parameters representing the properties of the entire modulator. However, some plots do utilize `instantaneous' modal parameters to simplify the comparison to results obtained from mode simulations.

Fig. \ref{fig:spectra} shows basic measurements of phase modulation measured from the MEMS modulator integrated with ring resonators and MZIs; as voltage is applied to the modulator, the aforementioned index and group index modulation modifies each structure's transmission spectrum. In these measurements, the voltage applied to the structures is swept from 0 V to 100 V, and then back down again to 0 V to examine the hysteresis characteristics of the actuators. For clarity, however, in \ref{fig:spectra}a,d,g we plot only the first half of these measurements. Additionally, the devices here were measured over a wide 1480-1630 nm bandwidth, but we show only a limited wavelength range of the ring resonator spectra so the resonances are clearly visible.

Figs. \ref{fig:spectra}a-c show results measured using an A-type modulator (depicted in cartoon at top) inside an add-drop ring resonator. In Fig. \ref{fig:spectra}a, the resonance shifts at higher voltages are clear, as are changes in extinction ratio due to insertion loss in the pull-in state. The zoomed in plots in Fig. \ref{fig:spectra}b more clearly show the behavior of the shifting resonance with voltage, with visible asymmetry between the increasing and decreasing voltage sweeps. When plotted in \ref{fig:spectra}c, this hysteresis is more clear. Such hysteresis is a well-known property of electrostatic MEMS modulators \cite{fedder1994simulation, seok2016large}. Also evident is the sudden pull-in effect at approximately 40 V in the increasing voltage sweep. Fine tuning of the phase modulation at voltages above the pull-in voltage is possible due to the slab curvature modulation discussed previously (Fig. \ref{fig:overview}c).

Figs. \ref{fig:spectra}d-f show similar results, but measured using a longer interaction length C-type modulator. Here, it is immediately evident in Fig.\ref{fig:spectra}d that the extinction ratio is nearly invariant with voltage, suggesting the insertion loss of this modulator is much lower than that of a). Due to the substantially longer interaction length, a phase shift of more than $2\pi$ is observed, so careful identification of resonances (Fig. \ref{fig:spectra}e) is critical to obtain accurate phase shift measurements. Additionally, the phase versus voltage curve shown in Fig. \ref{fig:spectra}f exhibits much more slight hysteresis than that of Fig. \ref{fig:spectra}c, but this behavior varies substantially across various devices, and is not a general advantage of the longer C-type structure.

We can also measure phase modulation using an MZI (Figs. \ref{fig:spectra}g-h). Here, we plot the full 1480-1630 nm bandwidth, illustrating the broadband nature of the modulator. Note that the clear envelope visible as an upper bound of the spectra in Fig. \ref{fig:spectra}g is due to the finite bandwidth of the grating couplers used for measurement and is not a limitation of the modulator. The phase versus voltage response in Fig. \ref{fig:spectra}h follows yet another distinct curve, illustrating the substantial device-to-device variation present in the actuators in this fabrication. Furthermore, the consistently deep extinction ratio of the fringes illustrates the very low insertion loss of the modulator.

\begin{figure*}
    \centering
    \includegraphics{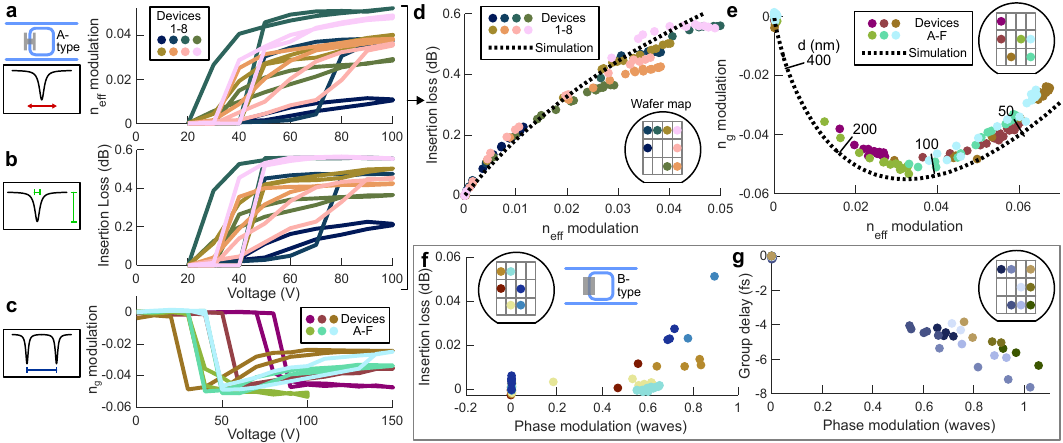}
    \caption{Phase, loss, and group delay modulation characteristics of several devices across a 4" wafer extracted from ring resonator spectra. a) Effective index modulation of extracted from resonance wavelength shifts, b) loss modulation extracted from changes in resonance depth (ER) and width (FWHM), and c) group index modulation extracted from the spacing between resonances (FSR). d) By plotting insertion loss versus effective index modulation, the dependence of the trace on the MEMS actuator characteristics can be eliminated and compared to simulation. e) Group index modulation plotted versus effective index modulation. f) When the same measurements are performed on long interaction length structures, the loss versus phase modulation for long interaction length devices shows very low insertion loss. g) Long interaction length structures still exhibit the predicted negative group delay modulation.}
    \label{fig:loss_and_dispersion}
\end{figure*}

In Fig. \ref{fig:loss_and_dispersion}, we move away from raw data measured on individual devices and onto extracted parameters measured for many (nominally) identical devices across a wafer. Note that for the phase and loss modulation results in Fig. \ref{fig:loss_and_dispersion}, the parameters are extracted in parallel for numerous resonances in add-drop ring resonators, then averaged together to produce final, low noise results. This averaging is performed over a wavelength range of 1545-1595 nm, so these results are compared to simulations performed at 1570 nm. Additionally, though we use a previously demonstrated method \cite{fedorov2025unambiguous} to extract loss, in this work there are special considerations which are discussed in Supplemental Section S4.

Figs. \ref{fig:loss_and_dispersion}a-e show results from A-type devices, with integrated parameters converted to instantaneous parameters under the assumption of a uniform interaction length $L$ = 13 \textmu m. Fig. \ref{fig:loss_and_dispersion}a plots the effective index modulation (extracted from resonance wavelength shift) versus voltage for 8 devices across a wafer with identical designs. The actuators have substantially different performance from device-to-device, preventing simple interpretation of the results when plotted in this way. Similarly, when the insertion loss (extracted from resonance ER and FWHM) is plotted as a function of voltage (Fig. \ref{fig:loss_and_dispersion}b), substantial device-to-device variation is also visible. Notably, however, the phase modulation and loss modulation curves for corresponding devices in Fig. \ref{fig:loss_and_dispersion}a and b (curves of the same color across the two different plots) have visually similar shapes, suggesting correlation. When insertion loss is plotted as a function of effective index modulation (Fig. \ref{fig:loss_and_dispersion}d), the device-to-device variation is nearly eliminated (with the exception of deviations at high effective index modulations, which are discussed further below). The existence of a consistent response across all devices when only optical properties are plotted supports the hypothesis that the device-to-device variation is indeed due to the MEMS actuators here and not inconsistency in the fundamental optical effect. In other words, this plot shows that the only parameter that varies significantly from device to device is the voltage versus $d$ relationship, not the $d$ versus $n_\text{eff}$ or $d$ versus loss relationship. If either of the latter relationships were device-dependent, the $n_\text{eff}$ versus loss relationship would also be device dependent, which is observed not to be the case.

When data can be expressed in terms of these purely (spatially instantaneous) optical parameters, it can easily be compared with simple eigenmode simulations. Here, we approximate insertion loss from simulations by calculating from the mode mismatch loss between the unmodulated mode profile and the modulated mode profile (e.g., Figs. \ref{fig:big_picture}d and e), then squaring it, as there are two facets at which the propagating mode experiences loss. The data in  Fig. \ref{fig:loss_and_dispersion}d has excellent agreement with these simulated results, never deviating by more than 0.05 dB for low levels of modulation. At very high modulations, the loss deviates slightly more substantially, up to 0.1 dB below the simulated curve. We believe this may be due to the high voltages beginning to distort the MEMS slab so substantially that curvature cannot be neglected, even within the short 13 \textmu m physical length of the A-type structures. This curvature would yield larger values of $d$ at the edges of the interaction length (where the mode mismatch is experienced) than at the center, allowing for greater modulation for a given loss than the flat slab approximation predicts.

Fig. \ref{fig:loss_and_dispersion}c shows the measured group index change. This parameter is extracted from the FSR between adjacent ring resonances. It is important to note here that since the interaction length of the short devices is only 13 \textmu m, the changes in FSR are extremely small. In particular, we only see changes in FSR on the order of 1 pm, which is below the open-loop wavelength precision specifications of our laser (5 pm). As such, we use a fiber interferometer in our setup to perform dynamic calibration of our wavelength, which improves the precision of our wavelength measurements by an order of magnitude (less than 0.5 pm). This, along with other noise-reducing considerations were required to obtain the results shown here -- for more details refer to Supplemental Section S5. As with Figs. \ref{fig:loss_and_dispersion}a-b, the device-to-device variation in MEMS actuator performance still results in difficult to interpret results in Fig. \ref{fig:loss_and_dispersion}c.

Fortunately, when we plot group index modulation versus effective index modulation (Fig. \ref{fig:loss_and_dispersion}e), we clearly observe a local minimum in the anomalously dispersive response, which is one of the key predictions of the theory presented earlier in this work (Fig. \ref{fig:big_picture}c). Furthermore, there is once again excellent agreement between the experimental data and results from simple mode simulations. In this plot we also label the simulation curve with markers indicating the MEMS-waveguide distance $d$ at various points. These distance markers illustrate the extremely fine, nm-scale control of the distance $d$ enabled by the slab curvature modulation technique. The slight deviation between the experimental and simulation curve (less than 0.005 refractive index units) could be due to slight differences in the material index and layer thicknesses in the simulation compared to the fabricated devices, or due to curvature of the MEMS slab within the interaction length.

While the lossy behavior illustrated in d) agrees well with simulation, which validates our modeling, this loss is certainly not ideal for most practical applications. As previously mentioned, the slab curvature of long interaction length devices is designed to eliminate this loss. In Fig. \ref{fig:loss_and_dispersion}f, we test this hypothesis by plotting the insertion loss versus phase modulation extracted from B-type devices, which have an 87 \textmu m physical length. Indeed, these devices are capable of $2\pi$ modulation while keeping insertion loss below 0.05 dB. Many devices achieve insertion losses below 0.02 dB, which is comparable with the uncertainty of the ring resonator method used to measure this loss. In Supplemental Section S4 we also compare the total round-trip loss of these finished devices to that measured during inline testing performed prior to MEMS fabrication. This comparison shows that the excess loss introduced by MEMS integration is also extremely low, typically adding less than 0.02 dB of round-trip loss in these ring resonator structures. Finally, Fig. \ref{fig:loss_and_dispersion}g presents the group delay modulation of these structures. The data points in this plot are more scattered than those in \ref{fig:loss_and_dispersion}e, which is very likely from device-to-device variations in slab curvature. One trend noticeable in Fig. \ref{fig:loss_and_dispersion}g is that devices on the east side of the wafer exhibit a lower magnitude of group delay modulation (at a given level of phase modulation) than those further west on the wafer. In Supplemental Section S6 we plot the impact of slab curvature on the group delay modulation characteristics. This cross-wafer trend is consistent with a higher curvature of the devices on the west side of the wafer, which could be due to nonuniform strain in the MEMS slab or variations in the thickness of the bumpers.

\section{Discussion and Conclusion}

Despite over 40 years of prior research into evanescent MEMS modulators, to the best of our knowledge the insights demonstrated here enabling the control of dispersion and loss are novel, and have substantial practical importance for use of MEMS modulators in various applications. Utilization of the very large effective index (and group index modulations) possible with evanescent MEMS modulators has previously been limited by the large mode mismatch-induced insertion losses at each end of the modulator \cite{hah2011mechanically}. The use of slab curvature here to overcome this loss mechanism enables evanescent MEMS modulators to realize very high modulations in a short interaction length with little excess loss, a capability not accessible to nearly any other modulator platform. Nuanced control of slab curvature also has several possibilities for higher dimensional control of the modulator's properties beyond what is shown in this work. In Supplemental Section S6 we show that an actuator with control over both the modulator's curvature and minimum position above the waveguide $d_\text{min}$ can access regions of the plot in Fig. \ref{fig:loss_and_dispersion}g that do not just fall on a single line, but over a finite area. This capability is true not just for the group delay modulation, but also for the GVD modulation. We believe this would represent a form of broadband dispersion reconfigurability that is not currently possible with any other on-chip modulator platform.

Despite the promises of this technology, the demonstration here has several clear limitations. Primarily, the actuators here are merely an early design built with the purposes of demonstrating the optical principles of interest in this work. During measurements, on long interaction length structures we observed slow response times and drifts on the order of seconds during actuation. However, these issues are well-known to occur in MEMS systems before optimization of the actuator implementation, and could be due to numerous factors that are not controlled for in these experiments (e.g., the disconnected `floating' electrodes in the electrostatic actuator interfering with operation, or uncontrolled humidity and pressure in the testing environment). We also note here that limitations and inconsistencies in the actuator performance in fact provided some utility in producing the experimental results in shown Figs. \ref{fig:loss_and_dispersion}d-g. Due to the pull-in effect of the actuators, ideal actuators would not be able to stably access actuator positions with $d>90$ nm, as this is the height of the bumpers used here, which defines $d$ in the pull-in state. It is expected during design that when the voltage is decreased below the level required to hold the MEMS slab tightly on the bumpers, the slab will pull-up, being fully removed from the waveguide evanescent field. However, when measurements were performed on fabricated devices (data used in Figs. \ref{fig:spectra} and  \ref{fig:loss_and_dispersion}a-e), we frequently observed phase and loss modulations consistent with a wide range of $d$ values greater than 90 nm. This ability to access intermediate positions was useful in this work, however many applications require only access of the `up' state and slight adjustments in the pull-in state. Applications which require more continuous control of the actuator position over a wide range would require a substantially different actuator design.

One final important limitation of MEMS technologies in general is the substantial increase in fabrication complexity over other platforms such as thermo-optic modulators. Even so, our MEMS fabrication process has several advantages over many other approaches; one major general advantage of evanescent MEMS devices is that no movable waveguide structures are required. Indeed, all MEMS fabrication here was patterned using low resolution UV lithography, which substantially simplifies fabrication and is more conducive for mass manufacturing. The fabrication performed here is not dissimilar to previous works which demonstrated devices manufactured on large glass substrates \cite{hong2023large} - fabrication on such large substrates introduces numerous possibilities for large scale integrated photonics. Finally, even considering the high fabrication complexity of MEMS devices, many of the capabilities proposed here (such as power-efficient modulation at visible wavelengths and high dimensional dispersion control) are unique qualities of MEMS modulators simply not possessed by thermo-optic modulators.

There are numerous applications in which the unique properties of these modulators could be utilized. To begin, in interferometric switches, to obtain large extinction ratios with low insertion loss, the modulator must exhibit a wavelength independent $\pi$ phase shift over the operating bandwidth with a very tight tolerance. As is evident in Eq. \ref{eq:phase_neff}, a single value of $\Delta n_\text{eff}$ is only capable of achieving this precise $\pi$ shift over a small range of wavelengths. Given this, anomalously dispersive modulation (in which the effective index modulation is increasing with wavelength) is required to obtain $\pi$ phase shifts over broad bandwidths. It can be shown that wavelength independent (to the first order) phase shifts are obtained precisely when $\Delta n_g = 0$. This regime is accessible with the devices shown here, making them uniquely capable of unusually broadband interferometric switches. While there are various other proposed broadband integrated photonic switches based on MEMS \cite{seok2016large, akihama2011ultra, han2015large}, to our knowledge all such platforms require suspended waveguide structures.

Another useful application of the group index modulation properties in these switches is for controllable integrated photonic delay lines. It is well-known that the resolution of on-chip Fourier transform spectrometers is related to the maximum achievable group delay \cite{kita2018high, johnson2025interferometric}. As such, the large group index modulations achieved here may allow for high resolution on-chip spectrometers with low power consumption that can operate over a wide range of wavelengths, due to the wide transparency window of the silicon nitride waveguide. 

Beyond switches and delay lines, there are numerous other interesting ways in which the unique properties of these modulators may enable on-chip systems extremely challenging to realize with existing modulator technology. For one, dispersion control is central to technologies that facilitate ultrafast pulse shaping and phase matching of nonlinear processes \cite{armstrong1962interactions, weiner2011ultrafast}. The dispersive properties of these modulators can be envisioned as useful for these purposes even as demonstrated here (with the MEMS devices placed above a staight waveguide), given much longer interaction lengths. On the other hand, shorter interaction lengths would be sufficient if these devices were integrated on top of Bragg reflectors. Dynamic control of the MEMS position and curvature over the length of such reflectors would allow for reconfigurability of their wavelength-dependent amplitude and phase response at a microsecond timescale, an unprecedented capability.

In summary, in this work we demonstrated novel dispersive properties of evanescent MEMS modulators and illustrated how MEMS slab curvature can be engineered to be advantageous for control of the loss and dispersion alike. These modulators exhibit unique capabilities that have practical relevance for a variety of applications. We believe that integration of these MEMS devices into a variety of integrated photonic systems in the future will create numerous new opportunities for the use of integrated photonics in previously untapped wavelength ranges and applications.

\section{Acknowledgments}
This work was supported by the National Science Foundation (ECCS-2023730, ECCS- 494 2217453) and the National Aeronautics and Space Administration (80NSSC24K1476, 80NSSC21K0799, 80NSSC23CA128).

K.J. thanks Vladimir Fedorov for his valuable contributions to the automated testing setup used to characterize the devices in this study.

\bibliography{main}

\end{document}